\documentclass[11pt]{article}
\usepackage{osid}
\usepackage{amsmath}
\usepackage{amssymb}
\newcommand{\we}{\wedge}
\newcommand{\ot}{\otimes}
\newcommand{\ti}{\times}

\newcommand{\pa}{\partial}
\newcommand{\cD}{{\cal D}}

\newcommand{\cJ}{{\cal J}}
\newcommand{\cS}{{\cal S}}

\newcommand{\cF}{{\cal F}}
\newcommand{\cO}{{\cal O}}
\newcommand{\cP}{{\cal P}}
\newcommand{\cR}{{\cal R}}

\newcommand{\cE}{{\cal E}}

\newcommand{\cH}{{\cal H}}

\newcommand{\la}{\langle}
\newcommand{\ra}{\rangle}

\newcommand{\raa}{\rightarrow}

\newcommand{\R}{{\mathbf R}}
\newcommand{\C}{{\mathbf C}}
\newcommand{\D}{\text{d}}
\newcommand{\wh}{\widehat}
\newcommand{\wt}{\widetilde}

\newcommand{\ep}{\rule{1.5mm}{2.5mm}\\}
\newcommand{\bA}{{\bf A}}
\newcommand{\bB}{{\bf B}}

\newcommand{\tr}{\mbox{$\text{Tr}$}}
\newcommand{\bra}[1]{\ensuremath{\langle #1 |}}
\newcommand{\ket}[1]{\ensuremath{| #1\rangle}}

\newcommand{\kb}[2]{\ensuremath{| #1\rangle\!\langle #2 |}}
\mathchardef\za="710B  %\alpha
\mathchardef\zb="710C  %\beta
\mathchardef\zg="710D  %\gamma
\mathchardef\zd="710E  %\delta
\mathchardef\zve="710F %\epsilon
\mathchardef\zz="7110  %\zeta
\mathchardef\zh="7111  %\eta
\mathchardef\zvy="7112 %\theta
\mathchardef\zi="7113  %\iota
\mathchardef\zk="7114  %\kappa
\mathchardef\zl="7115  %\lambda
\mathchardef\zm="7116  %\mu
\mathchardef\zn="7117  %\nu
\mathchardef\zx="7118  %\xi
\mathchardef\zp="7119  %\pi
\mathchardef\zr="711A  %\rho
\mathchardef\zs="711B  %\sigma
\mathchardef\zt="711C  %\tau
\mathchardef\zu="711D  %\upsilon
\mathchardef\zvf="711E %\phi
\mathchardef\zq="711F  %\chi
\mathchardef\zc="7120  %\psi
\mathchardef\zw="7121  %\omega
\mathchardef\ze="7122  %\varepsilon
\mathchardef\zy="7123  %\vartheta
\mathchardef\zf="7124  %\varomega
\mathchardef\zvr="7125 %\varrho
\mathchardef\zvs="7126 %\varsigma
\mathchardef\zf="7127  %\varphi
\mathchardef\zG="7000  %\Gamma
\mathchardef\zD="7001  %\Delta
\mathchardef\zY="7002  %\Theta
\mathchardef\zL="7003  %\Lambda
\mathchardef\zX="7004  %\Xi
\mathchardef\zP="7005  %\Pi
\mathchardef\zS="7006  %\Sigma
\mathchardef\zU="7007  %\Upsilon
\mathchardef\zF="7008  %\Phi
\mathchardef\zW="700A  %\Omega

\newcommand{\be}{\begin{equation}}
\newcommand{\ee}{\end{equation}}

\newcommand{\bea}{\begin{eqnarray}}
\newcommand{\eea}{\end{eqnarray}}
\newcommand{\beas}{\begin{eqnarray*}}
\newcommand{\eeas}{\end{eqnarray*}}

\newtheorem{prop}{Proposition}

\newtheorem{cor}{Corollary}

\title{Symmetries, group actions, and entanglement}

\author{Janusz Grabowski \\{\footnotesize\it Polish Academy of Sciences, Institute of Mathematics,
\'Sniadeckich 8, P.O. Box 21, 00-956 Warsaw, Poland \& E-mail jagrab@impan.gov.pl}\\[2ex]
Marek Ku\'s \\{\footnotesize\it Center for Theoretical Physics, Polish Academy of Sciences,
Aleja Lotnik{\'o}w 32/46, 02-668 Warszawa, Poland \& E-mail marek.kus@cft.edu.pl}\\[2ex]
Giuseppe Marmo \\{\footnotesize\it Dipartimento di Scienze Fisiche, Universit\`{a}
``Federico II'' di Napoli and Istituto Nazionale di Fisica Nucleare, Sezione di Napoli, Complesso
Universitario di Monte Sant Angelo, Via Cintia, I-80126 Napoli, Italy \& E-mail marmo@na.infn.it}}

\begin{document}
\maketitle

\begin{abstract}
We address several problems concerning the geometry of the space
of Hermitian operators on a finite-dimensional Hilbert space, in
particular the geometry of the space of density states and
canonical group actions on it. For quantum composite systems we
discuss and give examples of measures of entanglement.
\end{abstract}

\section{Introduction}
\label{sec:intro} In his book \cite{Di}, Dirac uses the
description of the interference phenomena, via the superposition
rule, to justify the requirement of linearity on the carrier space
of states to deal with quantum evolution. However, the
probabilistic interpretation of state vectors  forces on us the
identification of physical (pure) states with points of  the
complex projective space associated with the starting vector space
of "states". With this identification a (global) linear structure
is no more available, now interference phenomena will be described
with the help of a connection (Pancharatnam connection)
\cite{pancharatnam}. The Hermitian structure available on the starting
Hilbert space of "states" induces a K\"ahlerian structure on the
complex projective space. The induced action of the unitary group,
projected from the one on the Hilbert space, allows for the
imbedding of the complex projective space into the dual of the Lie
algebra of the unitary group itself by means of the momentum map
associated with the symplectic  action of the group. Within this
ambient space, by means of the available linear structure, it is
possible to construct convex combinations of pure states (rank-one
projectors) and build the totality of density states. The space so
constructed is not linear and gives rise to interesting
geometrical structures. To deal with these various
non-linearities, recently \cite{GKM}, elaborating on previous
geometrical approaches to quantum mechanics \cite{all}, we have
considered the differential geometry of density states. This
approach seems to be quite appropriate to deal with composite
systems and the set of separable and entangled states which  do
not carry a linear structure. In this note we would like to
further elaborate on some subtle points which we have encountered
in our previous paper \cite{GKM}. To make the paper self-contained
we briefly recall the main results from our previous treatment.
The paper is organized as follows: after introducing notations and
conventions, in Section 3 we describe the basic geometric
structures useful in description of the density states, in
particular the invariant K\"ahler structures on the orbits of
unitary representations as well as an action of the general linear
group on the dual of the Lie algebra of the unitary group. The
description of this action in terms of the Kraus operators along
with some of their properties is further elaborated in Sections 5
and 6, where a general linear group action on density states is
discovered. In Section 7 we describe the geometry of the set of
density states as a convex body, in particular we discuss the
smoothness of its boundary. The rest of the paper is devoted to
description of the composite systems: in Section 8 we describe the
pure and mixed states of such systems in terms of the Segre
imbedding and give a general prescription for the construction of
entanglement measures. Examples of such constructions are given in
Sections 9 and 10 for bipartite and multipartite systems.

\section{Notations and conventions}
\label{sec:notations} Let $\cH$ be an $n$-dimensional Hilbert
space with the Hermitian product $\la x,y\ra_\cH$ being, by
convention, $\C$-linear with respect to $y$ and anti-linear with
respect to $x$. The unitary group $U(\cH)$ acts on $\cH$
preserving the Hermitian product and it consists of those complex
linear operators $A\in gl(\cH)$ on $\cH$ which satisfy
$AA^\dag=I$, where $A^\dag$ is the Hermitian conjugate of $A$,
i.e., $ \la Ax,y\ra_\cH=\la x,A^\dag y\ra_\cH$. The geometric
approach to Quantum Mechanics is based on the realification
$\cH_\R$ of $\cH$ considered as a K\"ahler manifold
$(\cH_\R,J,g,\zw)$ with canonical structures: a complex structure
$J$, a Riemannian metric $g$, and a symplectic form $\zw$. The
latter come from the real and the imaginary parts of the Hermitian
product, respectively. After the obvious identification of the
vectors tangent to $\cH_\R$ with $\cH_\R$, all these structures
are constant (do not depend on the actual point of $\cH_\R$) and
read
$$J(x)=i\cdot x,\qquad g(x,y)+i\cdot\zw(x,y)=\la x,y\ra_\cH.$$
We have obvious identities
$$J^2=-I,\quad \zw(x,Jy)=g(x,y),\quad g(Jx,Jy)=g(x,y),\quad
\zw(Jx,Jy)=\zw(x,y).$$

Fixing an orthonormal basis $(e_k)$ of $\cH$ allows us to identify
the Hermitian product $\la x,y\ra_\cH$ on $\cH$ with the canonical
Hermitian product on $\C^n$ of the form $\la
a,b\ra_{\C^n}=\sum_{k=1}^n\overline{a_k}{b_k}$, the group $U(\cH)$
of unitary transformations of $\cH$ with $U(n)$, its Lie algebra
$u(\cH)$ with $u(n)$, etc. In this picture
$(a_{jk})^\dag=(\overline{a_{kj}})$ and $(T^\dag
T)_{jk}=\la\za_j,\za_k\ra$, where $\za_k=(t_{jk})\in\C^n$ are
columns of the matrix $T=(t_{jk})$. The choice of the basis
induces (global) coordinates $(q_k,p_k)$, $k=1,\dots,n$, on
$\cH_\R$ by
$$\la e_k,x\ra_\cH=(q_k+i\cdot p_k)(x),$$
in which $\pa_{q_k}$ is represented by $e_k$ and $\pa_{p_k}$ by
$i\cdot e_k$. Hence the complex structure reads
$$J=\sum_k\left(\pa_{p_k}\ot\D q_k-\pa_{q_k}\ot\D p_k\right),$$
the Riemannian tensor
$$g=\sum_k\left(\D q_k\ot\D q_k+\D p_k\ot\D p_k\right)=\frac{1}{2}
\sum_k\left(\D q_k\vee\D g_k+\D p_k\vee\D p_k\right)$$ and the
symplectic form
$$\zw=\sum_k\D q_k\we \D p_k,$$
where $x\vee y=x\ot y+y\ot x$ is the symmetric, and $x\we y=x\ot
y-y\ot x$ is the wedge product. In complex coordinates
$z_k=q_k+i\cdot p_k$ one can write the Hermitian product as the
complex tensor $\la\cdot,\cdot\ra_\cH=\sum_k\D\overline{z}_k\ot \D
z_k$.

\medskip
One important convention we want to introduce is that we will
identify the space of Hermitian operators $A=A^\dag$ with the dual
$u^*(\cH)$ of the (real) Lie algebra $u(\cH)$, according to the
pairing between Hermitian $A\in u^*(\cH)$ and anti-Hermitian $T\in
u(\cH)$ operators $\la A,T\ra=\frac{i}{2}\cdot\text{Tr}(A T)$. The
multiplication by $i$ establishes further a vector space
isomorphism $u(\cH)\ni T\mapsto iT\in u^*(\cH)$ which identifies
the adjoint and the coadjoint action of the group $U(\cH)$,
$\text{Ad}_U(T)=U T U^\dag$. Under this isomorphism $u^*(\cH)$
becomes a Lie algebra with the Lie bracket
$[A,B]=\frac{1}{i}[A,B]_-$ (where $[A,B]_-=AB-BA$ is the
commutator bracket), equipped additionally with the scalar product
\be\label{metric1}\la A,B\ra_{u^*}=\frac{1}{2}\text{Tr}(AB)
\ee
and an additional algebraic operation, the Jordan product (or
bracket) $[A,B]_+=AB+BA$. The scalar product is invariant with
respect to both: the Lie bracket and the Jordan product
\bea\label{invariant1}
\la[A,\zx],B\ra_{u^*(\cH)}&=&\la A,[\zx,B]\ra_{u^*(\cH)},\\
\label{invariant2} \la[A,\zx]_+,B\ra_{u^*(\cH)}&=&\la
A,[\zx,B]_+\ra_{u^*(\cH)}.
\eea
and it identifies once more $u^*(\cH)$ with its dual, $u^*(\cH)\ni
A\mapsto\widehat{A}=\frac{1}{i}A\in u(\cH)$, so vectors with
covectors. Under this identification the metric (\ref{metric1})
corresponds to the invariant metric
\be\label{metric2}\la
\wh{A},\wh{B}\ra_{u}=\frac{1}{2}\text{Tr}(AB)
\ee
on $u(\cH)$ which can be viewed also as a contravariant metric on
$u^*(\cH)$.

\smallskip \noindent
For a (real) smooth function $f$ on $\cH_\R$ let us denote by
$grad_f$ and $Ham_f$ the gradient and the Hamiltonian vector field
associated with $f$ and the Riemannian and the symplectic tensor,
respectively. In other words, $g(\cdot,grad_f)=\D f$ and
$\zw(\cdot,{Ham}_f)=\D f$. We can define also the corresponding
Poisson and Riemann-Jordan brackets of smooth functions on $\cH$
by
$$\{ f,f'\}_\zw=\zw(Ham_f,Ham_{f'}),\quad \{ f,f'\}_g=g(grad_f,grad_{f'})
$$
and the total bracket by $\{ f,f'\}_\cH=\{ f,f'\}_g+i\{
f,f'\}_\zw$. Note that any complex linear operator $A\in gl(\cH)$
induces a linear vector field $\widetilde{A}$ on $\cH$ by
$\widetilde{A}(x)=Ax$ and a quadratic function
$f_A(x)=\frac{1}{2}\la x,Ax\ra_\cH$. The function $f_A$ is real if
and only if $A$ is Hermitian, $A=A^\dag$. It is easy to see the
following.
\begin{theorem}{Theorem}
\begin{description} \item{(a)} For Hermitian $A$ we have
$$grad_{f_A}=\widetilde{A}\quad \text{and}\quad
Ham_{f_A}=\widetilde{i\,A}.$$ \item{(b)} For all $A,B\in gl(\cH)$
we have $\{ f_A,f_B\}_\cH=f_{2AB}$. In particular,
\bea\label{br1}
\{ f_A,f_B\}_g&=&f_{AB+BA},\\
\label{br2}\{ f_A,f_B\}_\zw&=&f_{-i(AB-BA)}. \eea
\end{description}
\end{theorem}

\medskip\noindent
The unitary action of $U(\cH)$ on $\cH$ is in particular
Hamiltonian and induces a momentum map $\zm:\cH_\R\raa u^*(\cH)$.
The fundamental vector field associated with $\frac{1}{i}A\in
u(\cH)$, where $A\in u^*(\cH)$ is Hermitian, reads
$\widetilde{iA}$, since
$$\frac{\D}{\D t}_{\mid_{t=0}}\exp{(-\frac{t}{i}A)}(x)=iA(x).$$ The
Hamiltonian of the vector field $\widetilde {iA}$ is $f_A$, so the
momentum map is defined by
$$\la\zm(x),\frac{1}{i}A\ra=f_A(x)=\frac{1}{2}\la x,Ax\ra_\cH.$$
But by our convention
$$\la\zm(x),\frac{1}{i}A\ra=\frac{i}{2}\text{Tr}(\zm(x)\frac{1}{i}A)=
\frac{1}{2}\text{Tr}(\zm(x)A),$$ so that $\text{Tr}(\zm(x)A)=\la
x,Ax\ra_\cH$ and finally, in the Dirac notation,
\be\zm(x)=\mid x\ra\la x\!\mid.
\ee
Hence, the image of this momentum map consists of all non-negative
Hermitian operators of rank $\le 1$. The space of all
non-negatively defined operators, i.e.\ of those $\zr\in gl(\cH)$
which can be written in the form $\zr=T^\dag T$ for a certain
$T\in gl(\cH)$, we denote by $\cP(\cH)$. It is a convex cone in
$u^*(\cH)$. The set of density states $\cD(\cH)$ is distinguished
in the cone $\cP(\cH)$ by the equation $\text{Tr}(\zr)=1$, so it
is a convex body in $u^*(\cH)$. Denote by $\cP^k(\cH)$ (resp.,
$\cD^k(\cH)$) the set of all non-negative hermitian operators
(resp., density states) of rank $k$. In the standard terminology,
$\cD^1(\cH)$ is the space of pure states, i.e.\ the set of
one-dimensional orthogonal projectors $\mid x\ra\la x\mid$, $\Vert
x\Vert=1$.

It is known that the set of extreme points of $\cD(\cH)$ coincides
with the set $\cD^1(\cH)$ of pure states (see Corollary
\ref{extreme}). Hence every element of $\cD(\cH)$ is a convex
combination of points from $\cD^1(\cH)$. The space $\cD^1(\cH)$ of
pure states can be identified with the complex projective space
$P\cH\simeq\C P^{n-1}$ via the projection $$\cH\setminus\{ 0\}\ni
x\mapsto \frac{\mid\! x\ra\la x\!\!\mid}{\Vert
x\Vert^2}\in\cD^1(\cH)$$ which identifies the points of the orbits
of the $\C\setminus\{ 0\}$-group action by complex homoteties. It
is well known that $\cD^1(\cH)$ is canonically a K\"ahler
manifold. This will be the starting point for the study of
geometry of $u^*(\cH)$ and the set $\cD(\cH)$ of all density
states.

\section{Geometry of $u^*(\cH)$}
\label{sec:geometry1} Recall that $u^*(\cH)$ is canonically an
Euclidean space with the scalar product $\la
A,B\ra_{u^*}=\frac{1}{2}\text{Tr}(AB)$. We have also seen that
$u^*(\cH)$ is canonically a Lie and a Jordan algebra with the
brackets $[A,B]=\frac{1}{i}(AB-BA)$ and $[A,B]_+=AB+BA$,
respectively. Note also that, for $A$ being Hermitian, $f_A$ is
the pullback $f_A=\zm^*(\widehat{A})=\widehat{A}\circ\zm$, where
$\widehat{A}=\la A,\cdot\ra_{u^*}=\frac{1}{i}A\in u(\cH)$. The
linear functions $\widehat{A}$ generate the cotangent bundle
$\text{T}^*u^*(\cH)$, so that (\ref{br1}) and (\ref{br2}) mean
that the momentum map $\zm$ relates the contravariant analogs of
$g$ and $\omega$, respectively, with the linear contravariant
tensors $R$ and $\zL$ on $u^*(\cH)$ corresponding to the Jordan
and Lie bracket, respectively. The Riemann-Jordan tensor $R$,
defined in the obvious way,
\be
R(\zx)(\wh{A},\wh{B})=\la\zx,[A,B]_+\ra_{u^*}=\frac{1}{2}\text{Tr}
(\zx(AB+BA)),
\ee
is symmetric and the tensor
\be
\zL(\zx)(\wh{A},\wh{B})=\la\zx,[A,B]\ra_{u^*}=\frac{1}{2i}\text{Tr}
(\zx(AB-BA)),
\ee
is the canonical Kostant-Kirillov-Souriau Poisson tensor on
$u^*(\cH)$. They form together the complex tensor
\be(R+i\cdot\zL)(\zx)(\wh{A},\wh{B})=2\la\zx,AB\ra_{u^*}=
\text{Tr}(\zx AB)
\ee
and the momentum map relates this tensor with the Hermitian
product.

\medskip
\noindent On $u^*(\cH)$ consider the (generalized) distributions
$D_\zL$ and $D_R$ induced by the tensor fields $\zL$ and $R$,
respectively. To be more precise, Denote by $\wt{\cJ}$ and
$\wt{\cR}$ the $(1,1)$-tensors on $u^*(\cH)$, viewed as a vector
bundle morphism induced by the contravariant tensors $\zL$ and
$R$, respectively, $\wt{\cJ},\wt{R}:\text{T}u^*(\cH)\raa
\text{T}u^*(\cH)$, where
\beas
\wt{\cJ}_\zx(A)&=&[A,\zx]=\zL_\zx(A),\\
\wt{\cR}_\zx(A)&=&[A,\zx]_+=R_\zx(A),
\eeas
for $A\in u^*(\cH)\simeq\text{T}_\zx u^*(\cH)$. The image of
$\wt{\cJ}$ is $D_\zL$ and the image of $\wt{\cR}$ is $D_R$. It is
easy to see that the tensors $\wt{\cJ}$ and $\wt{\cR}$ commute and
\be\wt{\cJ}_\zx\circ\wt{\cR}_\zx(A)=
\wt{\cR}_\zx\circ\wt{\cJ}_\zx(A)=[A,\zx^2].
\ee
We will consider also the generalized distributions
$D_1=D_R+D_\zL$ and $D_0=D_R\bigcap D_\zL$.

\medskip\noindent
{\it 1. Distribution ${D_\zL}$.}

\medskip\noindent
It is well known that the distribution $D_\zL$ can be integrated
to a generalized foliation $\cF_\zL$ whose leaves are orbits of
the canonical $U(\cH)$-action on $u^*(\cH)$ given by $U(\cH)\ti
u^*(\cH)\ni(U,\zx)\mapsto U\zx U^\dag\in u^*(\cH)$. They are
represented by the spectrum of the operator, i.e.\ $\zr$ and $\zr'$
belong to the same $U(\cH)$-orbit if and only if they have the
same set of eigenvalues (counted with multiplicities). Moreover,
the orbits are symplectic leaves of the Poisson structure $\zL$
with the corresponding $U(\cH)$-invariant symplectic form
$\zh^\cO$. These symplectic structures can be extended to
canonical K\"ahler structures as shows the following.
\begin{theorem}{Theorem}
\begin{description}
\item{(a)} \ $\wt{\cJ}_\zx^2$ is a self-adjoint with respect to
$\la\cdot,\cdot\ra_{u^*}$ and non-positively defined operator on
$u^*(\cH)$ with the kernel $D_\zL(\zx)^\bot$.

\item{(b)} The $(1,1)$-tensor $\cJ$ on $u^*(\cH)$ defined by
\be\label{J}
\cJ_\zx(A)=\left\{
\begin{array}{ll}
0 & \mbox{if $A\in D_\zL(\zx)^\bot$}\\
\wt{\cJ}_\zx\circ\left(-(\wt{\cJ}_\zx)^2_{\mid
D_\zL(\zx)}\right)^{-\frac{1}{2}} (A) & \mbox{if $A\in
D_\zL(\zx)$}
\end{array}\right.
\ee
satisfies $\cJ^3=-\cJ$ and induces an $U(\cH)$-invariant complex
structure $\cJ$ on every $U(\cH)$-orbit ${\cO}$.

\item{(c)} For every $U(\cH)$-orbit ${\cO}$ the tensor
\be\zg^{{\cO}}_\zx(A,B)=\zh^{{\cO}}_\zx(A,\cJ_\zx(B))
\ee
is an $U(\cH)$-invariant Riemannian metric on ${\cO}$ and
\be\label{46}\zg^{{\cO}}_\zx(\cJ_\zx(A),B)=\zh^{{\cO}}_\zx(A,B),
\quad A,B\in D_\zL(\zx).
\ee
In particular, $({\cO},\cJ,\zh^\cO,\zg^\cO)$ is a homogeneous
K\"ahler manifold. Moreover, if $\zx\in u^*(\cH)$ is a projector,
$\zx^2=\zx$, and $\zx\in\cO$, then $\cJ_\zx=\wt{\cJ}_\zx$ and
$\zg^\cO(A,B)=\la A,B\ra_{u^*}$.
\end{description}
\end{theorem}
{\bf Remark.} The tensor $\cJ$ is canonically and globally
defined. It is however not smooth as a tensor field on $u^*(\cH)$.
It is smooth on the open-dense subset of regular elements and, of
course, on every $U(\cH)$-orbit separately.

\medskip\noindent
{\it 2. Distribution ${D_R}$.}

\medskip\noindent
We have some similar results for the tensor $\wt{\cR}$ which
however are not completely analogous, since the distribution $D_R$
is not integrable.
\begin{theorem}{Theorem}
\begin{description}
\item{(a)} $\wt{\cR}_\zx^2$ is a self-adjoint with respect to
$\la\cdot,\cdot\ra_{u^*}$ and non-negatively defined operator on
$u^*(\cH)$ with the kernel $D_\zL(\zx)^\bot$.

\item{(b)} The $(1,1)$-tensor $\cR$ on $u^*(\cH)$ defined by
\be\label{J1}
\cR_\zx(A)=\left\{
\begin{array}{ll}
0 & \mbox{if $A\in D_R(\zx)^\bot$}\\
\wt{\cR}_\zx\circ\vert(\wt{\cR}_\zx)_{\mid D_R(\zx)}\vert^{-1} (A)
& \mbox{if $A\in D_R(\zx)$}
\end{array}\right.
\ee
satisfies $\cR^3=\cR$.
\end{description}
\end{theorem}

\medskip\noindent
{\it 3. Distribution ${D_0}$.}

\medskip\noindent
The distribution $D_0=D_\zL\bigcap D_R$ can be described also as
the image of ${\cJ}_\zx\circ\cR_\zx=\cR_\zx\circ\cJ_\zx$. In other
words, $D_0(\zx)=\{[A,\zx^2]:A\in u^*(\cH)\}$.
\begin{theorem}{Theorem}
The distribution $D_0$ is integrable and the corresponding foliation $\cF_0$
is $U(\cH)$-invariant, $\cJ$-invariant and $\cR$-invariant, so
that $\cJ$ and $\cR$ induce on leaves of $\cF_0$ a complex and a
product structure, respectively. The leaves of the foliation
$\cF_0$ are also canonically symplectic manifolds with symplectic
structures being restrictions of symplectic structures on the
leaves of $\cF_\zL$, so the leaves of $\cF_0$ are K\"ahler
submanifolds of the $U(\cH)$-orbits in $u^*(\cH)$.
\end{theorem}
Note however, that $D_0$ coincides with $D_\zL$ on $\cP(\cH)$, so
that on density states the leaves of $\cF_0$ are just the orbits
of the unitary group action.

\medskip\noindent
{\it 4. Distribution ${D_1}$.}

\medskip\noindent
The distribution $D_1=D_\zL+D_R$ is the largest one carrying the
most qualitative information. It turns out to be related to a
$GL(\cH)$-action as shows the following.
\begin{theorem}{Theorem}\label{GL} \
\begin{description}
\item{(a)} The generalized distributions $D_1$ is involutive and
can be integrated to generalized foliations $\cF_{1}$ whose leaves
are the orbits of the $GL(\cH)$-action $GL(\cH)\ti
u^*(\cH)\ni(T,\zx)\mapsto T\zx T^\dag\in u^*(\cH)$.

\item{(b)} The Hermitian operators $\zr$ and $\zr'$ belong to the
same $GL(\cH)$-orbit if and only if they have the same number
$k_+$ of positive and the same number $k_-$ of negative
eigenvalues (counted with multiplicities). Such an orbit, denoted
by $u^*_{k_+,k_-}(\cH)$, is of (real) dimension $2nk-k^2$, where
$k=k_++k_-$, and its tangent spaces are described by the formula
\be\label{tangent}
B\in\text{T}_\zx u^*_{k_+,k_-}\Leftrightarrow \forall
x,y\in\text{Ker}(\zx )\ [\la Bx,y\ra_\cH=0].\ee

\item{(c)} Any $GL(\cH)$-orbit intersecting $\cP(\cH)$ lies
entirely in $\cP(\cH)$, so that $\cP(\cH)$ is stratified by the
$GL(\cH)$-orbits. The $GL(\cH)$-orbits in $\cP(\cH)$ are
determined by the rank of an operator, i.e.\ they are exactly
$\cP^k(\cH)$, $k=0,1,\dots,n$.
\end{description}
\end{theorem}
The proofs of all results in this section can be found in
\cite{GKM}.
\section{Explicit coordinates on $GL(\cH)$-orbits}
\label{sec:coordinates}
A choice of an orthonormal basis in $\cH$ gives an identification
$u^*(\cH)\simeq u^*(n)$. Let $J=\{ j_1,\dots,j_k\}\subset\{
1,\dots,n\}$. Denote $u^*_{J}(n)$ -- the open subset in the set
$u^*_{k}(n)$ of rank-$k$ Hermitian matrices consisting of matrices
$(a_{ij})\in u^*_{k}(n)$ for which the submatrix $(a_{rs})_{r,s\in
J}$ is invertible with the inverse $(a^{rs})_{r,s\in J}$.

\begin{theorem}{Lemma}
Let $\zx\in u^*_{J}(n)$. Then the matrix $\zx$ is uniquely
determined by its rows indexed by $ J$, according to the formula
$$  a_{ij}=\sum_{r,s\in J}{a_{ir}}a^{rs}\overline{a_{js}}.$$
\end{theorem}
\begin{theorem}{Theorem}
The maps
\beas & \Phi_J:u^*_{J}(n)\raa
\R^k\ti\C^{(2nk-k^2-k)/2}\simeq\R^{2nk-k^2},\\
& \Phi_J((a_{ij})_{i,j=1}^n)=((a_{ii})_{i\in J},(a_{rs})_
{r<s,s\in J})\eeas form a coordinate system in the manifold
$u^*_{k}(n)$.
\end{theorem}

\bigskip\noindent
{\bf Example.} For $ n=3$ and $ J=\{ 1,2\}$
$$\Phi_J\left(\begin{array}{ccc}d_1&a_1-ia_2&b_1-ib_2\\
a_1+ia_2&d_2&c_1-ic_2\\b_1+ib_2&c_1+ic_2&d_3\end{array}\right)
=(d_1,d_2,a_1,a_2)\in\R^4.$$ According to the above lemma, the
coordinates $(d_1,d_2,a_1,a_2)$ are sufficient to determine the
Hermitian matrix of rank 2 whose principal minor is non-vanishing.

\section{Kraus operators}
\label{sec:krauss1} The above defined $GL(\cH)$-action on
$u^*(\cH)$ is an action by means of particular {\it Kraus
operators} which, by definition, are operations $K_{\bf A}$ of the
form $\zr\mapsto\sum_iA_i\zr A_i^\dag$ for certain ${\bf
A}=(A_i)\in gl(\cH)^m$. It is well known that, considered as
operations on $gl(\cH)$, they are exactly the linear operations
preserving the Hermicity and exactly the linear operations which
are {\it completely positive} (cf. \cite{krauss}). Note that we do
not assume any normalization condition for $K_{\bf A}$, like
$\sum_iA_i^\dag A_i=I$. It is easy to see that the composition of
Kraus operators is again a Kraus operator, so that we can speak
of the semigroup of Kraus operators. Indeed, $K_\bA\circ K_{\bf
B}=K_{\bA\cdot{\bf B}}$, where $K_{\bA\cdot{\bf
B}}(\zr)=\sum_{i,j}(A_iB_j)\zr(A_iB_j)^\dag$.

Let us note that the space $gl(\cH)$ of all complex linear
operators on $\cH$ is canonically a Hilbert space with the
Hermitian product $\la A,B\ra_{gl}=\frac{1}{2}\text{Tr}(A^\dag
B)$. Using the Jamio\l kowski isomorphism one can identify the
Kraus operator $K_{\bf A}$ with a Hermitian operator on the
Hilbert space $gl(\cH)$, defined as ${\bf P}_{\bf A}=\sum_i
p_{A_i}$, where $p_{A_i}={\mid}A_i\ra\la A_i\!\mid$. Using the
spectral decomposition one can easily see that any Kraus operator
$K_\bA$ can be written in a canonical form $K_\bA(\zr)=K_{\bf
C}(\zr)=\sum_kC\zr C^\dag$, where the operators $C_k\in gl(\cH)$
are pairwise orthogonal, $\la C_k,C_{k'}\ra_{gl}=0$ for $k\ne k'$,
and that in this case $A_i=\za_{ik}C_k$ with
$\sum_i\za_{ik}\overline{\za_{ik'}}=\zd_k^{k'}$.

It is interesting that the operators of the $GL(\cH)$-action form
exactly the largest subgroup in the semigroup of Kraus operators.
Since in the literature we could find the analogous fact only for
the unitary group and the semigroup of normalized Kraus
operators, we will give a short proof of this general result.
\begin{theorem}{Theorem}
If a Kraus operator $K_{\bA}$ is invertible inside
Kraus operators, then $K_\bA(\zr)=A\zr A^\dag$ for certain $A\in
GL(\cH)$.
\end{theorem}
{\it Proof.} Assume $K_\bA^{-1}=K_{\bf B}$, so that
$\sum_{i,j}(A_iB_j)\zr(A_iB_j)^\dag=\zr$. According to the last
observation, $A_iB_j=\za_{ij}I$ for certain $\za_{ij}\in\C$ with
$\sum_{ij}\vert\za_{ij}\vert^2=1$. There is $\za_{i_0j_0}\ne 0$,
so $A_{i_0}$ and $B_{j_0}$ are invertible. Moreover $A_{i_0}$ is
proportional to $B_{j_0}^{-1}$, namely
$A_{i_0}=\za_{i_0j_0}B_{j_0}^{-1}$. Since we can clearly assume
that all $A_i$ and all $B_j$ are non-zero, we get that all $A_i$
and all $B_j$ are invertible. Indeed, $A_i$ is not invertible
implies that $A_iB_{j_0}=\za_{ij_0}I$ is not invertible thus zero,
so $A_i=0$, because $B_{j_0}$ is invertible, and similarly for
$B_j$. Moreover, every $A_i$ is proportional to $B_{j_0}^{-1}$,
whence to $A_{i_0}$, $A_i=\zg_iA_{i_0}$, $\zg_i\ne 0$. We get
therefore
$$K_\bA(\zr)=\sum_iA_i\zr A_i^\dag=(\sqrt{\zg}A_{i_0})\zr
(\sqrt{\zg}A_{i_0})^\dag,$$ where $\zg=\sum_i\vert \zg_i\vert^2$.
\ep

\section{Kraus and $GL(\cH)$-action on density states}
\label{sec:krauss2} We know already that $\cP^k(\cH)$ are
$GL(\cH)$-orbits in $u^*(\cH)$, so smooth submanifolds. We used
this fact in \cite{GKM} to show that their intersections with the
hyperplane $\text{Tr}(\zr)=1$, i.e.\ $D^k(\cH)$ are smooth
submanifolds too. Here we want to stress that, in fact, $D^k(\cH)$
can be regarded again as $GL(\cH)$-orbits with respect to an
action of $GL(\cH)$ on $\cD(\cH)$. Of course, we cannot apply
directly the action on $u^*(\cH)$, as $\cD(\cH)$ is not an
invariant set under this action. We can, however, modify the
$GL(\cH)$-action (in fact, even the Kraus action) on $\cP(\cH)$
in such a way that we get an action on density states whose orbits
are $D^k(\cH)$.

For, suppose $K_\bA(\zr)=\sum_iA_i\zr A_i^\dag$ is a {\it
non-degenerate} Kraus operation, i.e.\ $\sum_i A^\dag_iA_i\in
GL(\cH)$. Now, let us define an operation $\wt{K}_\bA$ on
$\cD(\cH)$ by
$$\wt{K}_\bA(\zr)=\frac{K_\bA(\zr)}{\text{Tr}(K_\bA(\zr))}.$$
The definition makes sense, since $\text{Tr}(K_\bA(\zr))>0$ for
any density state $\zr$. Indeed, since
$$\text{Tr}(K_\bA(\zr))=\text{Tr}(\sum_iA_i\zr A_i^\dag)=\text{Tr}
((\sum_iA_i^\dag A_i)\zr)$$ and $T=\sum_iA_i^\dag A_i$ is
invertible Hermitian and non-negative, so strictly positive, and
since $\zr$ in Hermitian non-negative, $\text{Tr}(T\zr)\le 0$ only
if $\zr=0$.

Now, it is a fundamental observation that we get in this way
really an action of the semigroup of Kraus operations, i.e.\ that
$\wt{K}_\bA\circ\wt{K}_\bB=\wt{K}_{\bA\cdot\bB}$. But it is
straightforward, since
$$\wt{K}_\bA\circ\wt{K}_\bB(\zr)=\frac{K_\bA\left(\frac{K_\bB(\zr)}
{\text{Tr}(K_\bB(\zr))}\right)}{\text{Tr}\left(K_\bA\left(\frac{K_\bB(\zr)}
{\text{Tr}(K_\bB(\zr))}\right)\right)}= \frac{K_\bA\circ
K_\bB(\zr)} {\text{Tr}(K_\bA\circ
K_\bB(\zr))}=\wt{K}_{\bA\cdot\bB}(\zr).$$ Note that, though this
action is not affine, convex sets are mapped into convex sets, as
\be\label{hull}\wt{K}_\bA(\zl\zr+(1-\zl)\zr')=\wt{\zl}\wt{K}_\bA(\zr)+(1-\wt{\zl})
\wt{K}_\bA(\zr'),\ee where
$$\wt{\zl}=\frac{\zl\text{Tr}(\wt{K}_\bA(\zr))}{\zl\text{Tr}(\wt{K}_\bA(\zr))
+(1-\zl)\text{Tr}(\wt{K}_\bA(\zr'))}.$$ Of course, we can reduce
this action of the semigroup of non-degenerate Kraus operations
to its largest subgroup, i.e.\ to $GL(\cH)$, obtaining the action
\be\label{action} GL(\cH)\ti\cD(\cH)\ni(A,\zr)\mapsto\wt{A}(\zr)=\frac{A\zr
A^\dag}{\text{Tr}(A\zr A^\dag)}\in\cD(\cH).\ee This action
preserves the rank and one can easily derive from Theorem \ref{GL}
the following.
\begin{theorem}{Theorem}
The decomposition of the convex body of density states $\cD(\cH)$
into orbits of the $GL(\cH)$-action (\ref{action}) is exactly the
stratification $\cD(\cH)=\bigcup_{k=1}^n\cD^k(\cH)$ into states of
a given rank.
\end{theorem}

\section{The geometry of density states}
\label{sec:geometry2}
The boundary $\pa\cD(\cH)$ of the convex body of density states
consists of the states of rank$<n$,
$\pa\cD(\cH)=\bigcup_{k=1}^{n-1}\cD^k(\cH)$. Each stratum
$\cD^k(\cH)$ is a smooth submanifold in $u^*(\cH)$. However, the
boundary $\pa\cD(\cH)$ is not smooth (except for the case $n=2$),
since its maximal stratum $\cD^{n-1}(\cH)$ is sewed up along
$\bigcup_{k=1}^{n-2}\cD^k(\cH)$ with edges there, as shows the
following (\cite{GKM}, Theorem 2).
\begin{theorem}{Theorem}\label{t3}
Every smooth curve $\zg:\R\raa u^*(\cH)$ through the convex body of density
states is at every point tangent to the stratum to which it actually belongs,
i.e.\ $\zg(t)\in\cD^k(\cH)$ implies \ ${\rm T}{\zg}(t)\in{\rm
T}_{\zg(t)}\cD^k(\cH)$.
\end{theorem}
The above theorem means that inside $\cD(\cH)$ we cannot smoothly
cross the stratum $\cD^k(\cH)$ of the boundary transversally to
it, like living on a cube we cannot smoothly cross an edge of the
cube tranversally to it.
\begin{cor} The boundary of the
convex body $\cD(\cH)$ of density states is a smooth submanifold
of $u^*(\cH)$ if and only if $\text{dim}(\cH)\le 2$.
\end{cor}

\medskip\noindent {\bf Remark.} It is well known that for $n=2$ the
convex set of density states is the three-dimensional ball and its
boundary --  the two-dimensional sphere (so called {\it Bloch
sphere}), so it is a smooth manifold.

\medskip\noindent
The next problem concerning the geometry of density states we will
consider is the question of the faces of $\cD(\cH)$, i.e.\ the
intersections of $\cD(\cH)$ with supporting affine hyperplanes. In
other words, a non-empty closed convex subset $K_0$ of a closed
convex set $K$ is called a {\it face} (or extremal subset) of $K$
if any closed segment in $K$ with an interior point in $K_0$ lies
entirely in $K_0$; a point $x$ is called an {\it extremal} point
of $K$ if the set $\{ x\}$ is a face of $K$. For $\zr\in\cD(\cH)$
consider the decomposition
\be\label{face}\cH=\cH_+^\zr\oplus\cH_0^\zr=\text{Im}(\zr)\oplus\text{Ker}(\zr),
\quad x=x^\zr_++x^\zr_0,\ee into the kernel and the image of
$\zr$.
\begin{theorem}{Theorem}
The face of $\cD(\cH)$ through $\zr\in\cD^k(\cH)$
consists of operators $A\in\cD(\cH)$ which, according to the
decomposition (\ref{face}), have the form $A(x^\zr_++x^\zr_0)=
A^\zr_+(x^\zr_+)$ for certain $A^\zr_+\in\cD(\text{Im}(\zr))$, so
it is affinely equivalent to the convex body of density states in
dimension $k$.
\end{theorem}
\begin{cor}\label{extreme}
Extremal points of $\cD(\cH)$ are exactly pure states.
\end{cor}
\begin{cor}
All non-trivial faces of $\cD(\cH)$ of maximal dimension, i.e.\ faces through
$\cD^{n-1}(\cH)$, are tangent to the sphere
$S(I/n;r)$, centered at $I/n$ with the radius
$r=\frac{1}{\sqrt{n(n-1)}}$, at points which are collinear with
the center and one of the pure states.
\end{cor}

\section{Composite systems and separability}
\label{sec:composite}
Suppose now that our Hilbert space has a fixed decomposition into the tensor
product of two Hilbert spaces $\cH=\cH^1\ot\cH^2$ (of dimensions $n_1$ and
$n_2$, respectively). This additional input is crucial in studying composite
quantum systems and it has a great impact on the geometrical structures we have
considered. The rest of this paper will be devoted to related problems.

Observe first that the tensor product map
\be\bigotimes:\cH^1\ti\cH^2\raa\cH=\cH^1\ot\cH^2
\ee
associates the product of rays with a ray, so it induces a
canonical imbedding on the level of complex projective spaces
\bea \text{Seg}:P\cH^1\ti P\cH^2&\raa& P\cH=P(\cH^1\ot\cH^2),\\
(\mid\!x^1\ra\la x^1\!\!\mid,\mid\!x^2\ra\la x^2\!\!\mid)
&\mapsto& \mid\!x^1\ot x^2\ra\la x^1\ot x^2\!\!\mid.
\eea
This imbedding of product of complex projective spaces into the
projective space of the tensor product is called in the literature
the {\it Segre imbedding} \cite{segre}. The elements of the image
$\text{Seg}(P\cH^1\ti P\cH^2)$ in $P\cH=P(\cH^1\ot\cH^2)$ are
called {\it separable} pure states (with respect to the
decomposition $\cH=\cH^1\ot\cH^2$).

The Segre imbedding is related to the (external) tensor product of
the basic representations of the unitary groups $U(\cH^1)$ and
$U(\cH^2)$, i.e.\ with the representation of the direct product
group in $\cH=\cH^1\ot\cH^2$,
\beas U(\cH^1)\ti U(\cH^2)\ni(\zr^1,\zr^2)&\mapsto& \zr^1\ot\zr^2\in
U(\cH)=U(\cH^1\ot\cH^2),\\
(\zr^1\ot\zr^2)(x^1\ot x^2)&=&\zr^1(x^1)\ot\zr^2(x^2).
\eeas
Note that $\zr^1\ot\zr^2$ is unitary, since the Hermitian product
in $\cH$ is related to the Hermitian products in $\cH^1$ and
$\cH^2$ by \be\label{sp1} \la x^1\ot x^2,y^1\ot y^2\ra_{\cH}=\la
x^1,y^1\ra_{\cH^1}\cdot \la x^2,y^2\ra_{\cH^2}.
\ee
The above group imbedding gives rise to the corresponding
imbedding of Lie algebras or, by our identification, of their
duals, which, with some abuse of notation, we will denote by
\be\label{imb}\text{Seg}:u^*(\cH^1)\ti u^*(\cH^2)\raa
u^*(\cH),\quad (\zx_1,\zx_2)\mapsto\zx_1\ot\zx_2.
\ee
The original Segre imbedding is just the latter map reduced to
pure states. In fact, a more general result holds true.
\begin{prop} The imbedding (\ref{imb}) maps
$\cD^k(\cH^1)\ti\cD^l(\cH^2)$ into $\cD^{kl}(\cH)$.
\end{prop}

\medskip\noindent
Let us denote the image $\text{Seg}(\cD^1(\cH^1)\ti\cD^1(\cH^2))$
--  the set of {\it separable pure states} -- by $\cS^1(\cH)$, and
its convex hull ${conv}\left(\cS^1(\cH)\right)$ -- the set of all
{\it separable states} in $u^*(\cH)$ -- by $\cS(\cH)$. The states
from
$$\cE(\cH)=\cD(\cH)\setminus\cS(\cH),$$
i.e.\ those which are not separable, are called {\it entangled
states}. It is well known (see e.g. \cite{GKM}) that $\cS^1(\cH)$
is exactly the set of extremal points of $\cS(\cH)$.

\medskip
Of course, by means of the tensor product of representations the
group product $GL(\cH^1)\ti GL(\cH^2)$ is canonically embedded in
$GL(\cH)$ like in the case of the unitary groups. The canonical
actions (\ref{action}) on $\cD(\cH^1)$ and $\cD(\cH^2)$ give rise
to the action to the corresponding action of $GL(\cH^1)\ti
GL(\cH^2)$ on $\cD(\cH^1)\ti\cD(\cH^2)$:
$$\wt{(A_1,A_2)}(\zr_1,\zr_2)=(\wt{A_1}(\zr_1),\wt{A_2}(\zr_2)).$$
On the other hand, $GL(\cH^1)\ti GL(\cH^2)$ as being embedded in
$GL(\cH)$ acts on $\cD(\cH)$.
\begin{theorem}{Theorem}
The aforementioned actions of $GL(\cH^1)\ti GL(\cH^2)$
are equivariant with respect to the Segre map
$$\text{Seg}:\cD(\cH^1)\ti\cD(\cH^2)\raa\cD(\cH).$$
Moreover, the set $\cS^1(\cH)$ of pure separable states and the
set $\cS(\cH)$ of all separable states are invariant with respect
to the canonical $GL(\cH^1)\ti GL(\cH^2)$-action on $\cD(\cH)$:
$$\wt{(A_1,A_2)}(\zr)=\frac{(A_1\ot A_2)\zr(A_1\ot A_2)^\dag}
{\mathrm{Tr} ((A_1\ot A_2)\zr(A_1\ot A_2)^\dag)}.$$
\end{theorem}
{\it Proof.} We get easily
\beas\wt{(A_1,A_2)}(\zr_1\ot\zr_2)&=&\frac{(A_1\zr_1A_1^\dag)\ot(
A_2\zr_2A_2^\dag)} {\text{Tr}\left((A_1\zr_1A_1^\dag)\ot(
A_2\zr_2A_2^\dag)\right)} =\frac{(A_1\zr_1A_1^\dag)\ot(
A_2\zr_2A_2^\dag)} {\text{Tr}(A_1\zr_1A_1^\dag)\text{Tr}
(A_2\zr_2A_2^\dag)}\\
&=& \wt{A_1}(\zr_1)\ot\wt{A_2}(\zr_2)
\eeas
that proves equivariance and the invariance of $\cS^1(\cH)$. The
invariance of $\cS(\cH)$ is not automatic, like in the case of the
group $U(\cH^1)\ti U(\cH^2)$, since the $GL(\cH^1)\ti
GL(\cH^2)$-action is not affine. On the other hand, (\ref{hull})
implies that the convex hull thus separability is respected:
\beas\wt{(A_1,A_2)}(\zl\zr_1\ot\zr_2+\zl'\zr'_1\ot\zr'_2)&=&
\wt{\zl}\wt{(A_1,A_2)}(\zr_1\ot\zr_2)+ \wt{\zl}'\wt{(A_1,A_2)}
(\zr'_1\ot\zr'_2)\\
&=&\wt{\zl}\wt{A_1}(\zr_1)\ot\wt{A_2}(\zr_2)+ \wt{\zl}'\wt{A_1}
(\zr'_1)\ot\wt{A_2}(\zr'_2),
\eeas
where $\zl'=1-\zl$ and $\wt{\zl}'=1-\wt{\zl}$. \ep

\smallskip\noindent
It is a common opinion that the $U(\cH^1)\ti U(\cH^2)$-action, as
preserving separability, is crucial for understanding the
entanglement. We see, however, that the $GL(\cH^1)\ti
GL(\cH^2)$-action preserves the separability as well and, since
the orbits are larger, it carries more qualitative information,
thus information which is easier to handle. As an example,
consider the corresponding $U(\cH^1)\ti U(\cH^2)$ and
$GL(\cH^1)\ti GL(\cH^2)$ orbits inside pure states.
\begin{prop} With respect to the Schmidt decomposition
\be\label{schmidt}
|\Psi\ra=\sum_{k=1}^m\zl_k|\zf^1_k\ra\ot|\zf^2_k\ra\ee of the
unit vector $|\Psi\ra\in\cH$ representing a pure state, the
$U(\cH^1)\ti U(\cH^2)$-orbits in $\cD^1(\cH)$ are distinguished by
the sequence $\zl_1\ge\dots\zl_m>0$, while the $GL(\cH^1)\ti
GL(\cH^2)$-orbits are distinguished only by the Schmidt number
$m$.
\end{prop}
{\it Proof.-} It is clear that the form of the Schmidt
decomposition is preserved by the $U(\cH^1)\ti U(\cH^2)$-action.
On the other hand, if we have two such decompositions
$\sum_{k=1}^m\zl_k|\zf^1_k\ra\ot|\zf^2_k\ra$ and
$\sum_{k=1}^m\zl_k|\zh^1_k\ra\ot|\zh^2_k\ra$, then, since
$\zf^1_k$ are pairwise orthonormal, since $\zf^1_k$ are pairwise
orthonormal, etc., there are $U^i\in U(\cH^i)$, i=1,2, such that
$U^1(\zf^1_k)=\zh^1_k$ and $U^2(\zf^2_k)=\zh^2_k$, $k=1,\dots,m$,
so
$$(U^1\ot U^2)(\sum_{k=1}^m\zl_k|\zf^1_k\ra\ot|\zf^2_k\ra)=
\sum_{k=1}^m\zl_k|\zh^1_k\ra\ot|\zh^2_k\ra.$$ A similar reasoning
gives the description of $GL(\cH^1)\ti GL(\cH^2)$-orbits, if we
only take into account that the exact values of the coefficients
$\zl_k$ are irrelevant for this action, as the group does not
respect the length of the vector. \ep

\smallskip\noindent The entangled states play an important role in
quantum computing and one of main problems is to decide
effectively whether a given composite state is entangled or not.
An abstract measurement of entanglement can be based on the
following observation \cite{GKM} (see also Ref. \cite{Vidal}).

\smallskip
Let $E$ be the set of all extreme points of a compact convex set
$K$ in a finite-dimensional real vector space $V$ and let $E_0$ be
a compact subset of $E$ with the convex hull
$K_0=\text{conv}(E_0)\subset K$. For every non-negative function
$f:E\raa\R_+$  define its extension $f_K:K\raa\R_+$ by
\be\label{cf}f_K(x)=\inf_{x=\sum t_i\za_i}\sum t_if(\za_i),
\ee
where the {\it infimum} is taken with respect to all expressions
of $x$ in the form of convex combinations of points from $E$.
Recall that that, according to Krein-Milman theorem, $K$ is the
convex hull of its extreme points.
\begin{theorem}{Theorem}\label{hullt}
For every non-negative continuous function $f:E\raa\R_+$ which
vanishes exactly on $E_0$ the function $f_K$ is convex on $K$ and
vanishes exactly on $K_0$
\end{theorem}
\begin{cor}
Let $F:\cD^1(\cH^1\ot\cH^2)\raa\R_+$ be a continuous function
which vanishes exactly on $\cS^1(\cH^1\ot\cH^2)$. Then
$$\zm=F_{\cD(\cH^1\ot\cH^2)}:\cD(\cH^1\ot\cH^2)\raa\R_+$$
is a measure of entanglement, i.e.\ $\zm$ is convex and
$\zm(x)=0\Leftrightarrow x\in\cS(\cH)$. Moreover, if $f$ is taken
$U(\cH^1)\ti U(\cH^2)$-invariant (resp. $GL(\cH^1)\ti
GL(\cH^2)$-invariant), then $\zm$ is $U(\cH^1)\ti
U(\cH^2)$-invariant (resp. $GL(\cH^1)\ti GL(\cH^2))$-invariant.
\end{cor}
\section{Bipartite entanglement}
\label{sec:bi} One of measures constructed according to the above described
prescription is the concurrence introduced originally as an auxiliary quantity,
used to calculate so called entanglement of formation of $2\times 2$
systems~\cite{wot98}. For pure states (\ref{schmidt}) it is defined as
\begin{equation}\label{concurrence}
c\left(\Psi\right):=\sqrt{1-\tr_1^{\phantom{k}}\rho_1^2}=\sqrt{||\Psi||^2-
\tr_1^{\phantom{k}}\big(\tr_2^{\phantom{k}}(\kb{\Psi}{\Psi})
\cdot\tr_2^{\phantom{k}}(\kb{\Psi}{\Psi})\big)},
\end{equation}
where $\tr_i$, $i=1,2$, denotes tracing over the $i$-th subsystem,
and $\rho_1:=\tr_2\kb{\Psi}{\Psi}$. It is clear that, indeed, it
vanishes for separable states (for which
$\tr_1^{\phantom{k}}\rho_1^2=1$). For further generalizations (see
Section 10) it is convenient to rewrite (\ref{concurrence}) in
slightly different form. To this end let us define:
\begin{equation}\label{A}
A:{\cal H}\otimes{\cal H}\rightarrow {\cal H}\otimes{\cal H}, \quad
A=4P_-^{(1)}\otimes P_-^{(2)},
\end{equation}
where $P_-^{(i)}$ is the orthogonal projection on the antisymmetric part
$\cH^i\wedge\cH^i$ of the tensor product $\cH^i\otimes\cH^i$, and we identify
in an obvious manner $\cH^1\otimes\cH^2\otimes\cH^1\otimes\cH^2$ and
$\cH^1\otimes\cH^1\otimes\cH^2\otimes\cH^2$. It is now a matter of a
straightforward calculation that $c(\Psi)$ can be expressed as:
\begin{equation}\label{}
c\left(\Psi\right)=\sqrt{\big(\bra\Psi\otimes\bra\Psi\big)
A\big(\ket\Psi\otimes\ket\Psi\big)}.
\end{equation}

Extension of the concurrence (\ref{concurrence}) to mixed states
is defined {\it via} (\ref{cf}), i.e.\ \begin{equation}\label{concurrencemixed}
c(\rho)=\inf_{\rho\,=\sum t_i\kb{\Psi_i}{\Psi_i}}\sum t_ic(\Psi_i).
\end{equation}
Calculation of $c(\rho)$ for an arbitrary mixed state $\rho$ requires a high
dimensional optimization procedure, it is however possible to derive a lower
bound for it which in general suffices to discriminate between a separable and
an entangled state. Our bound is given by a purely algebraic expression easily
evaluated for arbitrary states and can be, if needed, tightened numerically by
optimizing over a parameter space of much lower dimensionality than it is
demanded by the original definition (\ref{concurrencemixed})
\cite{mkb:04,mckb:05}. To this end we first replace the $\ket{\Psi_i}$ by the
subnormalized states $\ket{\psi_i}=\sqrt{\,t_i}\,\ket{\Psi_i}$ in
Eq.~(\ref{concurrencemixed}). Given a valid decomposition $\rho =\sum
\kb{\phi_i}{\phi_i}$ into $M$ subnormalized states $\{ \ket{\phi_i},
i=1,\ldots,M \}$, any other suitable set $\{ \ket{\psi_i}, i=1,\ldots,N \}$
such that
\begin{equation}\label{subdecomp}
\rho =\sum \kb{\psi_i}{\psi_i},
\end{equation}
is obtained \cite{hugston} by:
\be \ket{\psi_i}=\sum_{j=1}^M V_{ij}
\ket{\phi_j}, \quad V\in{\mathbb C}^{N\times M}, \quad
\sum_{i=1}^N\overline{V}_{ik}V_{ij}=\delta_{jk},
\label{allens}
\ee
where both $N$ and $M$ are not smaller than the rank $r$ of $\rho$. It can be
shown \cite{uhl98} that for the purposes of the present considerations, it is
enough to take $N\le n_1^2n_2^2$. We can now choose as the starting point eg.\
the decomposition $\{\ket{\phi_i}, i=1,\ldots,r \}$ of $\rho$ defined in terms
of its (subnormalized) eigenvectors
\begin{equation}\label{eigrho}
\rho=\sum_{i=1}^r\kb{\phi_i}{\phi_i}, \quad
\rho\ket{\phi_i}=a_i\ket{\phi_i},
\end{equation}
where $a_i, i=1,\ldots,r$, are non-vanishing eigenvalues of
$\rho$. Now the concurrence can be rewritten as:
\begin{equation}\label{conc1}
c(\rho)=\inf_V\sum_i\sqrt{ \bigl[\big(V\otimes V\big) {\cal A}\big(
V^\dagger\otimes V^\dagger\big)\bigr]_{ii}^{ii}},
\end{equation}
where
\be
{\cal A}_{jk}^{lm}=
\big(\bra{\phi_l}\otimes\bra{\phi_m}\big) A
\big(\ket{\phi_j}\otimes\ket{\phi_k}\big),
\ee
and the {\it infimum} is now taken on over matrices $V$ fulfilling
(\ref{allens}). The expression simplifies further if expressed in terms of
eigenvectors of $A$, i.e.\ $ A=\sum_{\alpha=1}^m\kb{\chi_\alpha}{\chi_\alpha}$,
$\alpha=1,\hdots,m=n_1(n_1-1)n_2(n_2-1)/4$. Namely
\be
c(\rho)=\inf_V\sum_i\sqrt{ \sum_\alpha \Big|\left[VT^\alpha
V^T\right]_{ii}\Big|^2}, \label{concT}
\ee
where $T_{jk}^{\alpha}=\bra{\chi_{\alpha}}(\ket{\phi_j}\otimes\ket{\phi_k})$.
Obviously any given decomposition (\ref{subdecomp}) provides a straightforward
upper bound of the concurrence,
$c(\rho)\le\sum_i\sqrt{\sum_\alpha\left|T^\alpha_{ii}(\psi)\right|^2}$. From
the point of view of distinguishing separable and entangled states it is much
more interesting to find a lower bound for $c$ in an an effective way. To this
end let us write, using Cauchy-Schwarz inequality,
\begin{equation}\label{lb1}
\sqrt{ \sum_\alpha \Big|\left[VT^\alpha
V^T\right]_{ii}\Big|^2}\sqrt{\sum_\alpha\Big|z_\alpha|^2}\ge
\sum_\alpha\Big|\left[Vz_\alpha T^\alpha
V^T\right]_{ii}\Big|\ge\Big|\sum_\alpha\left[Vz_\alpha T^\alpha
V^T\right]_{ii}\Big|,
\end{equation}
for arbitrary $z_\alpha$, $\alpha=1,\ldots,m$.  We obtain thus:
\be c(\rho)\ge\inf_V \sum_{i=1}^N \Bigl|\Bigl[V \Bigl(\sum_\alpha
z_\alpha T^\alpha\Bigr)V^T \Bigl]_{ii}\Bigr|, \label{lb2} \ee for
arbitrary $z_\alpha$ such that $\sum_\alpha|z_\alpha|^2=1$. The
{\it infimum} over $V$ can be effectively performed and is given
by $\mathrm{max}\left\{\lambda_1-\sum_{i>1}\lambda_i,0\right\}$,
where $\lambda_j$ are the singular values of $T=\sum_\alpha
z_\alpha T^\alpha$, i.e.\ the square roots of the eigenvalues of
the positive hermitian matrix $TT^{\dagger}$ in the decreasing
order \cite{uhl00}. The obtained bound still depends on the choice
of the $z_\alpha$, what allows to tighten the estimate. Thus, one
is left with an optimization problem on an $2m$-dimensional
sphere. Note that the constraint $\sum_\alpha|z_\alpha|^2=1$ is by
far simpler to implement than $V^\dagger V$ (cf. \ref{allens}).
Moreover, the dimension $m$ of optimization space is significantly
reduced as compared to the dimension $n_1^3 n_2^3$ of the original
optimization problem defined by Eq.~(\ref{allens}). Let us,
however, point out that any choice of $z_\alpha$ gives some lower
bound and taking eg.\ all but one $z_\alpha$ equal to zero, we can
dispose of the optimization entirely if we are only interested
whether $c$ is positive, which is enough to establish
nonseperability of the state in question \cite{mkb:04}.

\section{Multipartite entanglement}
\label{sec:multi}

Separability of multipartite systems, where the Hilbert space of
the whole system $\cal H$ has a fixed decomposition into the
tensor product of Hilbert spaces ${\cal H}={\cal
H}^1\otimes\ldots,{\cal H}^K$ of subsystems of dimensions
$n_1,\ldots,n_K$ is defined by a straightforward extension of the
two-component case (cf.\ Section 8), i.e.\ via a canonical
imbedding of the product of projective spaces $P{\cal
H}^1\times\ldots\times P{\cal H}^K$ into the projectivisation of
the tensor product $P({\cal H}^1\otimes\ldots\otimes{\cal H}^K)$
and the corresponding imbedding on the level of Lie algebras and
their duals. The pure separable states are thus identified with
the image under this imbedding of ${\cal D}^1({\cal
H}^1)\times\ldots\times{\cal D}^1({\cal H}^K)$ and its convex hull
with the set of all separable states.

In order to investigate the separability of multipartite states we proposed the
following generalization of the concurrence considered in the preceding
section \cite{mkb:05}. Let us, in an analogy with (\ref{A}) define
\begin{equation}\label{multiA}
A_{\{s_j\}}:{\cal H}\otimes{\cal H}\rightarrow {\cal
H}\otimes{\cal H}, \quad A_{\{s_j\}}=2^K\bigotimes_{j=1}^K
P_{s_j}^{(j)},
\end{equation}
where $s_j=\pm$ and $P_-^{(j)}$ (respectively $P_+^{(j)}$) are
orthogonal projections on the antisymmetric (resp.\ symmetric)
subspace ${\cal H}^j\wedge{\cal H}^j$ (resp.\ ${\cal H}^j\vee{\cal
H}^j$), define concurrence for pure states as
\begin{equation}\label{multiC}
c_{\{s_j\}}\left(\Psi\right)=\sqrt{(\bra\Psi\otimes\bra\Psi)
A_{\{s_j\}}(\ket\Psi\otimes\ket\Psi)},
\end{equation}
and its extension to mixed states by
\begin{equation}\label{multimixedC}
c_{\{s_j\}}(\rho)=\inf_{\rho\,=\sum t_i\kb{\Psi_i}{\Psi_i}}\sum
t_ic_{\{s_j\}}(\Psi_i).
\end{equation}
Closer examination of the action of $A_{\{s_j\}}$ reveals that if
an odd number of projectors on anti-symmetric subspaces appears in
its definition, the corresponding $c_{\{s_j\}}\left(\Psi\right)$
vanishes identically. Moreover, if $s_i=+$ for all $i$, i.e. when
only projections on symmetric subspaces are involved,
$A_{\{s_j\}}$ is not helpful in detecting entanglement
\cite{mkb:05,dkmb}.

The techniques which were devised to ease the task of estimating the
concurrence for arbitrary states in the bipartite case in the previous section,
can be generalized in a straightforward manner, because the algebraic structure
of the above $K$-partite concurrences is strictly identical to the bipartite
definition (\ref{concurrencemixed}). Thus one can invoke the Cauchy-Schwarz and
the triangle inequality and bound the concurrence of an arbitrary mixed state
from below by
\begin{eqnarray}\label{multibound}
c_{\{s_j\}}(\rho)=\inf_V\sum_i\sqrt{\sum_\alpha\left|\left[VT^\alpha
V^T\right]_{ii}\right|^2}
&\ge&\inf_V\sum_i\left|\left[VTV^T\right]_{ii}\right|= \nonumber \\
&=&\mathrm{max}\Big\{\lambda_1-\sum_{j>1}\lambda_j,0\Big\},
\end{eqnarray}
where $T=\sum_\alpha z_\alpha T^\alpha$,
$T_{jk}^{\alpha}=\bra{\chi_{\alpha}}(\ket{\phi_j}\otimes\ket{\phi_k})$,
$A_{\{s_j\}}=\sum_{\alpha=1}\kb{\chi_\alpha}{\chi_\alpha}$, $\ket{\phi_k}$ are
the subnormalized eigenvectors of $\rho$, $V$ defines the transition between
different decompositions of $\rho$ as in (\ref{allens}), and $\lambda_j$ are
eigenvalues of $T^\dagger T$ in decreasing order. For the proof of the last
equality in (\ref{multibound}) see \cite{uhl00} or \cite{mckb:05}. As in the
bipartite case, the inequality (\ref{multibound}) holds for an arbitrary set of
complex numbers $z_\alpha$, such that $\sum_\alpha|z_\alpha|^2=1$, what allows
for further optimization.

The above does not only apply to the discrete set of concurrences discussed so
far, but also to the following continuous interpolation between them: Instead
of a single direct product of projectors onto symmetric and antisymmetric
subspaces, one may equally well consider convex combinations thereof,
\begin{equation}
\label{genA}
A=2^K\sum_{s_1,\dots,s_N} p_{s_1,\dots,s_K} P_{s_{1}}^{(1)}\otimes\ldots \otimes
P_{s_{K}}^{(K)}
\end{equation}
where $s_{i}\in\{+,-\}$, $p_{s_1,\dots,s_K}\geq 0$ and the summation is
restricted to contributions with an even, non-zero number of projectors onto
anti-symmetric subspaces. The corresponding pure-state concurrence
(\ref{multiC}) can be written in terms of the partial traces:
\begin{equation}
\label{eq:multitrace} c(\Psi) =\sqrt{ \sum\limits_{S \in 2^{\{1,\dots,K\}}}
\alpha_S \tr\left(\tr_A(\kb{\Psi}{\Psi})^2 \right)},
\end{equation}
where $2^{\{1,\dots,K\}}$ denotes the set of all subsets of $\{1,\dots,K\}$, and
\begin{equation}
\alpha_S=\sum\limits_{s_1,\dots,s_K} p_{s_1,\dots,s_K} \prod\limits_{i\in S}
s_i.
\end{equation}
Various choices of the coefficients $p_{s_1,\dots,s_K}$ allow to distinguish
different categories of multipartite entanglement. As an illustration, let us
focus on some exemplary tri- and four-partite concurrences.

The so called biseparable pure states (i.e.\ states taking the form
of tensor product of a state of one subsystem with a, possibly
entangled, state of the other two subsystems) in the tri-partite
case are easily detected with $A=P_+^{(1)}\otimes
P_-^{(2)}\otimes P_-^{(3)}$, $A=P_-^{(1)}\otimes
P_+^{(2)}\otimes P_-^{(3)}$, and $A=P_-^{(1)}\otimes
P_-^{(2)}\otimes P_+^{(3)}$. Whereas corresponding concurrences
vanish identically for bi-separable states like
$\ket{\psi}=\ket{\varphi_{12}}\otimes\ket{\zeta_3}$ for the first
and second choice of $A$, the last one which reduces to the
bi-partite concurrence of $\ket{\varphi_{12}}$.

Similarly, different kinds of separability are also captured in
larger systems. For example, concurrences defined with the help of
${\cal A}=4P_{s_1}^{(1)}\otimes P_{s_2}^{(2)}\otimes
P_{s_3}^{(3)}\otimes P_{s_4}^{(4)}$, $s_i=s_j=+$, and $s_k=-$ for
$i\neq k\neq j$, determine with respect to which bipartite
partition a mixed 4-particle state is separable, and quantify the
value of bi-, resp. tri-partite concurrences of the entangled
part \cite{mkb:05}.

\section{Acknowledgements}
This work was supported by the Polish Ministry of Scientific
Research and Information Technology under the (solicited) grant No
PBZ-Min-008/P03/03 and partially supported by PRIN SINTESI.

\end{document}